# A deep learning model for burn depth classification using ultrasound imaging

Sangrock Lee[1], Rahul[1*], James Lukan[2], Tatiana Boyko[2], Kateryna Zelenova[2], Basiel Makled[3], Conner Parsey[3], Jack Norfleet[3], and Suvranu De[1]

[1] Center for Modeling, Simulation and Imaging in Medicine, Rensselaer Polytechnic Institute, Troy, NY 12180, USA
[2] Department of Surgery, University at Buffalo-State University of New York, Buffalo, NY 14215, USA
[3] U.S. Army Futures Command, Combat Capabilities Development Command Soldier Center STTC, Orlando FL 32826, USA

**Abstract**

Identification of burn depth with sufficient accuracy is a challenging problem. This paper presents a deep convolutional neural network to classify burn depth based on altered tissue morphology of burned skin manifested as texture patterns in the ultrasound images. The network first learns a low-dimensional manifold of the unburned skin images using an encoder-decoder architecture that reconstructs it from ultrasound images of burned skin. The encoder is then re-trained to classify burn depths. The encoder-decoder network is trained using a dataset comprised of B-mode ultrasound images of unburned and burned *ex vivo* porcine skin samples. The classifier is developed using B-mode images of burned *in situ* skin samples obtained from freshly euthanized postmortem pigs. The performance metrics obtained from 20-fold cross-validation show that the model can identify deep-partial thickness burns, which is the most difficult to diagnose clinically, with 99% accuracy, 98% sensitivity, and 100% specificity. The diagnostic accuracy of the classifier is further illustrated by the high area under the curve values of 0.99 and 0.95, respectively, for the receiver operating characteristic and precision-recall curves. A *post hoc* explanation indicates that the classifier activates the discriminative textural features in the B-mode images for burn classification. The proposed model has the potential for clinical utility in assisting the clinical assessment of burn depths using a widely available clinical imaging device.

*Keywords:* Deep learning, Encoder-decoder CNN, Ultrasound imaging, Burn depth classification

---

[*] Corresponding author, Tel: +1 (518) 276-6351; Fax: +1 (518) 276-6025; Email: rahul@rpi.edu

1. **Introduction**

Burn care varies from supportive medical treatment to aggressive surgical intervention (Herndon, 2018). These diametrically different treatment strategies are dependent on the clinical evaluation of burn depth. However, the diagnostic accuracy of clinical assessment via visual and tactile inspection is only 50-80% (Ye and De, 2017) and lacks standardization. Burns traditionally are divided into three depths of tissue injury: epidermal, partial-thickness, and full-thickness (Jeschke et al., 2020; Ye and De, 2017). Partial-thickness burns are subdivided further into superficial-partial and deep-partial thickness burns. Early diagnosis of burn depth dictates the treatment planning, with superficial-partial thickness burns receiving medical management and deep-partial thickness burns often benefitting from early surgical excision and grafting (Jeschke et al., 2020). However, difficulty in assessing burn depth in the early period delays excision by days and remains a bottleneck in burn care (Monstrey et al., 2008). We propose a deep learning model for the objective evaluation of burn depth using ultrasound imaging modality.

While several noninvasive imaging techniques are proposed for objective evaluation of burn depths (Ye and De, 2017), laser-Doppler imaging (LDI) is the only standard alternative to clinical assessment. However, the diagnostic accuracy of LDI is less than 80% within the first 24-48 hours of burn injury (Hoeksema et al., 2009). Recently, state-of-the-art (SOTA) deep learning models have been employed to classify burn depths using digital images. A transfer learning technique based on a pre-trained residual neural network (ResNet101) is shown to classify burn depths using high-resolution digital images with average accuracy, sensitivity, and specificity of 0.91, 0.74, and 0.94, respectively (Cirillo et al., 2019). The pre-trained VGG16 (Simonyan and Zisserman, 2014) and ResNet50 (He et al., 2016) networks are shown to extract features from digital images of burn wounds that achieved maximum accuracy of 0.95 in classifying burn depths. However, these deep learning models lack an explanation of discriminative textural features that contributes to the burn depth classification. In addition, digital color images are susceptible to ambient illuminations that limit their clinical application.

Machine learning classification models differentiate between burnt human skin tissue and healthy tissue based on polarization-sensitive optical coherence tomography (OCT) images with average sensitivity, specificity, and accuracy of 0.92, 0.87, and 0.92, respectively (Dubey et al., 2018). A combination of features extracted from Raman spectroscopy measurements and OCT images of *ex vivo* porcine skin is shown to classify burn depths with an overall accuracy of 0.85 (Rangaraju et al., 2019). However, optical imaging modality has limitations in differentiating between burn depths due to the limitation on the penetration depth (Sen et al., 2016).

Unlike optical imaging modalities, ultrasonography can detect alterations in tissue acoustic impedance and therefore provides richer information regarding the penetration depth of a burn. Thermal treatment is known to alter the acoustic impedance of tissues (Fujii et al., 2012). The change in acoustic impedance is related to modifications in skin tissue morphology due to micro-detachment of the epidermis from the dermis and vacuolar cytoplasmic disintegration in the basal cell layer in case of partial-thickness burns, and denaturation of collagen and sometimes infiltration of the dermis by exudative cells in case of full-thickness burns (Moritz, 1947). The superposition of scattering acoustic echoes from the regions of contrasting acoustic impedance produces an intricate interference pattern that is manifested as speckles in the B-mode images (White, 2005). Our previous study shows an increase in speckle pattern for the thickness of the skin and subcutaneous tissues with increasing burn depth (Lee et al., 2020).

There have been several attempts to estimate burn depth using ultrasound signals (Ye and De, 2017). However, these studies rely on subjective evaluation of burn depths based on A- or B-mode signals and are often limited in accuracy (Ye and De, 2017). In recent work, we have shown that the engineered textural features obtained from B-mode images of *ex vivo* porcine skin can classify burn depths with an average accuracy of 93% (Lee et al., 2020). However, our preliminary study indicates that this approach yields an average accuracy of 88% in classifying burn depths in freshly euthanized pig *in situ* skin samples.

To overcome the limitations of existing models, we propose a deep convolutional neural network (CNN) architecture (BurnNet) to classify burn depths using B-mode ultrasound images. The BurnNet is designed to learn altered textural features, *i.e.*, speckle patterns, in the B-mode images, which are visually difficult to discern but critical for burn depth classification. BurnNet is developed in two stages. First, a source network learns a low-dimensional manifold of the unburned skin images using an encoder-decoder architecture that reconstructs it from the burned skin images. The source encoder-decoder network is trained using a dataset comprised of B-mode images of unburned and full-thickness burned *ex vivo* porcine skin samples. The encoder part is then repurposed as a target classifier and re-trained to classify burn depths. The target classifier is developed using B-mode images of burned *in situ* skin samples obtained from freshly euthanized postmortem pigs. The efficacy of the BurnNet is evaluated and benchmarked with the SOTA classifiers using several performance measures obtained from *k*-fold cross-validation. The *post hoc* explainability of the BurnNet is examined to test its ability to activate the discriminative textural features in the B-mode images of burned skin samples.

The rest of the article is organized as follows: Section 2 describes the dataset, the BurnNet architecture, training and validation approaches, and performance metrics used to evaluate and benchmark the BurnNet. The BurnNet performance metrics and benchmark studies are presented in Section 3. The *post hoc* explainability and trustworthiness of BurnNet are discussed in Section 4. This is followed by conclusions and future directions in Section 5.

## 2. Methods and materials

We first outline the data acquisition and processing steps. The architecture of the BurnNet and network training and validation approaches are described in the subsequent sections. Next, a brief description of the benchmark models and performance evaluation metrics is provided.

### 2.1. Data acquisition

***In situi*** **postmortem porcine skin dataset.** Controlled burning experiments were performed at the Erie County Medical Center in Buffalo, New York, on two postmortem pigs, within three hours of being euthanized. Each pig was placed on the surgical table in a prone position. A total of 20 locations on the dorsum of each pig were chosen to inflict a controlled burn. At each of these locations, one square inch area was shaved of hair and sterilized with chlorohexidine. A custom-developed burning device was used to produce the desired burn conditions. The burning device is a microprocessor-based temperature-controlled system consisting of a flexible silicone-rubber heating element (SRMU 100101, Omega Engineering Inc., CT; size: a square inch; wattage density: 10W/in$^2$; max exposure temperature: 450°F), a thermocouple for temperature measurement, a temperature controller (Arduino Uno R3, Arduino Inc.), and an electrically operated switch (relay). The flexible heater patch is instrumented with the thermocouple. The device is designed to maintain the desired temperature with minimal fluctuation. This type of device has been shown to inflict a controlled, reproducible, and homogeneous burn on the *in vivo* porcine skin (Krieger et al., 2005).

Four burn conditions corresponding to superficial-partial, deep-partial, and full-thickness burns were reproduced by varying exposer duration (at a fixed temperature) and heating temperature (for a given time period). It is well-known that burn depth is directly related to both the burn duration and burn temperature (Moritz and Henriques, 1947). Hence, we considered four different combinations of burn duration and burn temperature as surrogates of burn depth, based on literature (Abraham et al., 2015; Branski et al., 2008; Cuttle et al., 2006): (i) superficial-partial thickness burn (200ºF for 10s); (ii) deep-partial thickness burn (200ºF for 30s); (iii) light full-thickness burn (450ºF for 10s); and (iv) deep full-thickness burn (450ºF for 30s).

Ultrasound imaging was performed as soon as the skin temperature returned to room temperature, which was verified using an infrared camera (InfraCam, FLIR, OR). Ultrasound transmission gel was applied to the burn site, and a hand-held ultrasound probe (Ultrasonix L40-8/12 40-8MHz linear array probe with 10 MHz settings) was placed in contact with the skin. The image window size was set to $2.0 \times 1.3$ cm$^2$. The B-mode images were recorded on a SonixOne ultrasound unit. A total of 80 images, 20 each from four orientations, i.e., 0°, 45°, 90°, and 135° with respect to the spine, were collected for each of the four burn groups to ensure sufficient data sample size and eliminate the confounding effects due to the variability of tissue properties.

*Ex vivo* **porcine skin dataset.** The *ex vivo* porcine skin samples of unburned and full-thickness burned skin were prepared following the experimental protocol (Ye et al., 2018). The ultrasound imaging was performed using Ultrasonix L40-8/12 40-8MHz linear array probe with 10 MHz settings. The dataset comprised a total of 120 B-mode images – 60 each for the unburned and full-thickness burned skin samples.

## 2.2. Data processing

The B-mode ultrasound images were downsampled by a factor of 10, from the original size 338x213 pixels to the downsampled size of 34x22 pixels, to reduce the number of learnable BurnNet network parameters, primarily to address the small sample size, i.e., 80 B-mode images for each burn class. Fig. 1 compares a set of representative original B-mode images with the corresponding downsampled images of unburned skin and burned skin with four depths of burn injuries. From Fig. 1, we observe that the downsampled images retain the textural variations across the unburned and four burn groups. The training dataset was augmented by horizontally flipping images to create a diverse set of B-mode images for the efficient training of networks.

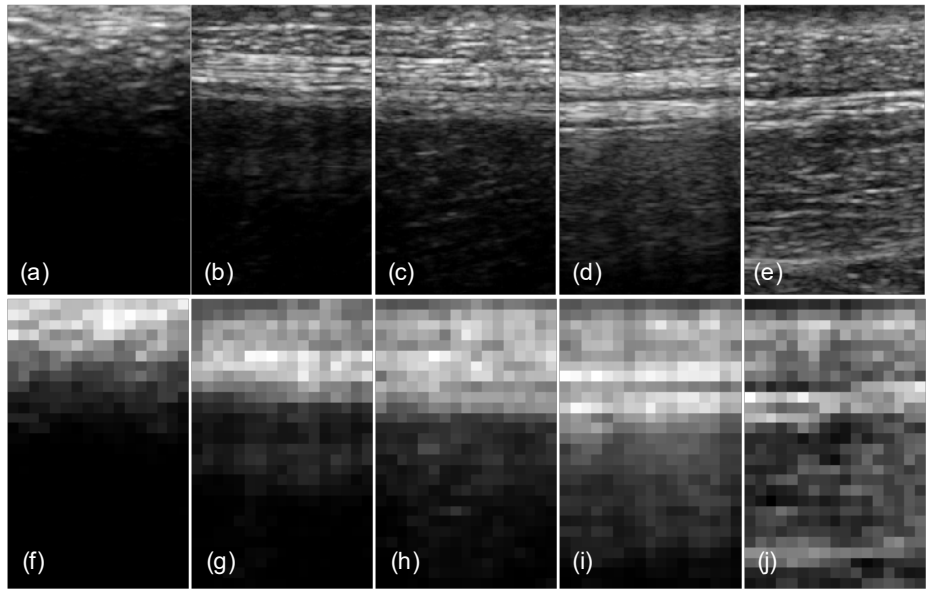

**Fig. 1.** Ultrasound B-mode images of *in situ* porcine skin: (a) unburned skin, (b)-(e) original images of skin with four burn depths, and (f)-(j) corresponding 10-factor downsampled images.

### 2.3. BurnNet architecture

The BurnNet is a transfer-learning-based CNN architecture. The source model of the BurnNet is a convolutional encoder-decoder network that maps B-mode ultrasound images of the burned skin to randomly chosen images of the unburned skin. From Fig. 1, the burned skin images show more speckle patterns than the unburned skin images. Hence, this mapping is analogous to denoising speckle patterns from the burned skin images to obtain unburned skin images. The network learns a low-dimensional manifold of the unburned skin images in the process. The denoising autoencoder architecture of the BurnNet makes it resilient to adversarial attacks, i.e., a small perturbation to the input image does not result in an incorrect prediction. The denoising autoencoder is known to prevent such adversarial attacks (Cho et al., 2020; Gan and Liu, 2020; Sahay et al., 2019). After training is complete, the pre-trained encoder is repurposed and re-trained as a target model to classify burn depths. The pre-training of the target classifier is intended to capture the main variations in the texture patterns of the B-mode images. The subsequent discriminative training further unfolds and separates the class structures, enhancing the classification accuracy.

A high-level overview of the BurnNet is shown in Fig. 2. The encoder is composed of four stacked encoding blocks and a bottleneck layer that serves to feature embedding. The kernel size for the convolutions and average pooling operations are set to 2x2 with stride 1. Notice that in the encoding blocks, average pooling layers, instead of max-pooling layers, are used to avoid the loss of textural information from the B-mode images. The decoder consists of four stacked decoding blocks. An additional deconvolution layer is added to the last decoding block to match the output image size. The kernel size for the convolution and deconvolution operations in the decoding layers are set to 2x2 with stride 2. The target classifier consists of the pre-trained encoder concatenated with a global average pooling layer and a fully connected layer with a single neuron (Fig. 2-bottom).

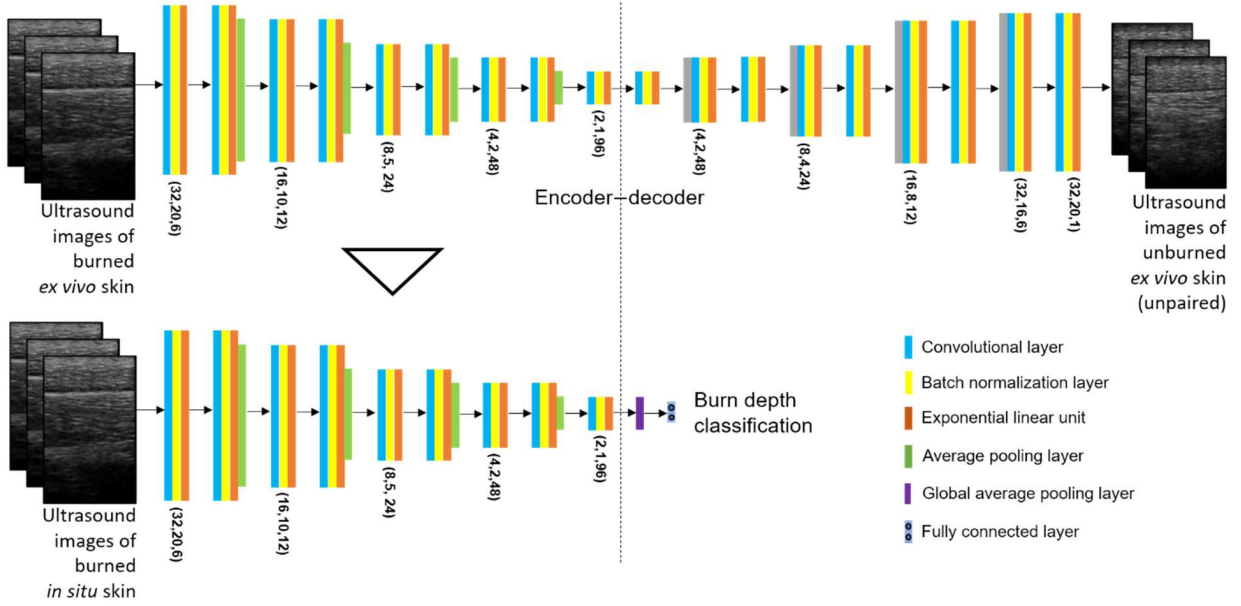

**Fig. 2.** A high-level schematic depiction of the BurnNet architecture. The **top** of the figure shows a source convolutional encoder-decoder with input burned and output unburned B-mode images of *ex vivo* porcine skin. The **bottom** of the figure shows the target classifier with input B-mode images of burned *in situ* porcine skin samples.

The encoder maps the pixel intensities of the burned skin image $\tilde{\mathbf{x}}$ to a hidden representation $\mathbf{y} = f_\theta(\tilde{\mathbf{x}})$, where $f_\theta$ is the encoding function. The decoder reconstructs the pixels of unburned skin image $\mathbf{z}$ from the hidden representation $\mathbf{y}$, *i.e.*, $\mathbf{z} = g_{\hat{\theta}}(\mathbf{y})$. The mapping $g_{\hat{\theta}}$ describes the decoding function. The source encoder-decoder network parameters are trained to minimize the reconstruction error $\mathcal{L}: \mathbb{R}^d \times \mathbb{R}^d \to \mathbb{R}$ over an empirical distribution $q(\tilde{\mathbf{x}}, \mathbf{x})$ of the training set, *i.e.*

$$\mathcal{J}(\theta, \hat{\theta}) = \mathbb{E}_q[\mathcal{L}(\mathbf{x}, (g_{\hat{\theta}} \circ f_\theta)(\tilde{\mathbf{x}}))], \qquad (1)$$

where $\mathbf{x}$ is the vector of pixel intensities of unburned skin image and $\mathbb{E}_q$ is the expected value operator with respect to the distribution $q(\tilde{\mathbf{x}}, \mathbf{x})$. The reconstruction error is given by the sum of the least absolute deviations ($l_1$-norm loss) and least square error ($l_2$-norm loss) between ground truth and predicted pixel intensities of unburned skin image, *i.e.*

$$\mathcal{L}(\mathbf{x}, \mathbf{z}) = \|\mathbf{x} - \mathbf{z}\|_1 + \|\mathbf{x} - \mathbf{z}\|_2^2. \qquad (2)$$

For the classification of burn depth, the decoder part of the source network is removed, and the encoder is repurposed as a target classifier described by $\mathbf{z}' = (h_\phi \circ f_\theta)(\tilde{\mathbf{x}})$. The parameters of the target classifier are trained to minimize the cross-entropy loss over the distribution $p(\tilde{\mathbf{x}})$ of the training set, *i.e.*

$$\mathcal{J}(\phi) = -\mathbb{E}_p[\mathbf{x}'\log(\mathbf{z}') + (1-\mathbf{x}')\log(1-\mathbf{z}')], \tag{3}$$

where $\mathbf{x}'$ is the ground truth burn depth class identifier, $\mathbf{z}'$ is the burn depth class predicted by the target classifier, and $\mathbb{E}_p$ is the expected value operator with respect to the distribution $p(\tilde{\mathbf{x}})$.

### 2.4. Training and validation

The BurnNet is trained in two steps. First, the source encoder-decoder network is trained in an unsupervised fashion to reconstruct B-mode images of unburned skin from full-thickness burned skin samples. The network is trained using *ex vivo* porcine skin dataset. The input burned and ground truth unburned skin images are not necessarily paired. Next, the target classifier with the pre-trained encoder from the source model is re-trained using B-mode images of *in situ* postmortem porcine skin dataset. The target model is trained for (i) a binary classification of deep-partial thickness burns vs. the rest of the burns depths and (ii) a multiclass classification of four burn groups. Both the source and target networks are trained using Adam optimizer via stochastic gradient descent on a minibatch. The learning rate and hyperparameters of the optimizer are set to default values given in (Kingma and Ba, 2014). The networks are trained for 2000 epochs with a minibatch of size 32. 20-fold cross-validation is performed for the independent assessment of the target classifier.

### 2.5. Benchmark models

The performance of the BurnNet is benchmarked against well-known classifiers for the binary classification of deep partial-thickness and the rest of the burn depths. The benchmark models include (i) linear discriminant analysis (LDA) (Mika et al., 1999), (ii) support vector machine (SVM) with radial basis function (RBF) kernel (Lee et al., 2020), and classifiers based on the state-of-the-art (SOTA) CNN such as (iii) VGG16 (Simonyan and Zisserman, 2014), (iv) ResNet (He et al., 2016), and (v) DenseNet (Huang et al., 2017). The fully connected layers of the SOTA classifiers are replaced by a global average pooling layer concatenated with a fully connected layer activated by a softmax function for classification. The same default setting of Adam optimizer is used to train the benchmark models.

The classifiers (i) and (ii) are trained using 19 textural features extracted from grey-level co-occurrence matrix (GLCM) (Haralick et al., 1973) derived from the original full resolution B-mode images. GLCM is a histogram of adjacent co-occurring pixel pairs of an image. Second-order statistical measures are calculated from a GLCM and used to characterize the specific texture features of an image. A list of textural features used in this study is provided in our previous work (Lee et al. 2020). Our previous study showed that the GLCM based textural features could differentiate between four burn depths with an average accuracy of 93% (Lee et al., 2020). This benchmark study is intended to evaluate the efficacy of engineered textural features in classifying burn depths in the *in situ* porcine skin samples. The classifiers (iii)-(v) inherit weights of the SOTA networks pre-trained on the ImageNet dataset. Only the weights of the fully connected layers are updated during the training process. These classifiers are trained using B-mode images from the *in situ* porcine skin dataset. The images are downsampled by a factor of 5, which is the minimum size of the images that the SOTA networks can process.

## 2.6. Model evaluation metrics

The performance of the classifiers is compared using receiver operating characteristic (ROC) curves, precision-recall (PR) curves, and several other performance metrics obtained from 20-fold cross-validation. The performance metrics for the binary classification include accuracy, sensitivity, specificity, F-score, and Matthew's correlation coefficient (MCC). In addition, the trustworthiness of BurnNet is benchmarked against SOTA classifiers described in Section 2.5. The trustworthiness of a model $M$ is measured using a set of recently proposed metrics such as question-answer trust, trust density, trust spectrum, and NetTrustScore (Wong et al., 2020).

The question-answer trust (Wong et al., 2020) for a given burned skin image $\tilde{\mathbf{x}}$ with burn depth class identifier $\mathbf{x}'$ is defined by

$$Q_{\mathbf{z}'}(\tilde{\mathbf{x}}, \mathbf{x}') = \begin{cases} C(\mathbf{x}'|\tilde{\mathbf{x}})^\alpha & \tilde{\mathbf{x}} \in R_{\mathbf{x}'=\mathbf{z}'|M} \\ (1 - C(\mathbf{x}'|\tilde{\mathbf{x}}))^\beta & \tilde{\mathbf{x}} \in R_{\mathbf{x}'\neq\mathbf{z}'|M} \end{cases}, \quad (4)$$

where $C(\mathbf{x}'|\tilde{\mathbf{x}})$ is the confidence of model $M$ in predicting $\mathbf{x}'$ for a given $\tilde{\mathbf{x}}$; $R_{\mathbf{x}'\neq\mathbf{z}'|M}$ is the space where $\mathbf{x}'$ and the predicted burn depth identifier $\mathbf{z}'$ do not match and $R_{\mathbf{x}'=\mathbf{z}'|M}$ is the space where $\mathbf{z}'$ is correctly identified as $\mathbf{x}'$; $\alpha$ and $\beta$ are the hyperparameters associated with reward for well-placed confidence and penalty for undeserved overconfidence, respectively. In this work, $\alpha = 1$ and $\beta = 1$ are set to reward and penalize equally. The confidence is given by the softmax outputs associated with the predicted burn depth class identifier $\mathbf{z}'$.

Trust density is the probability density distribution of question-answer trust (Wong et al., 2020) for all input burned skin images. We use a non-parametric density estimation with Gaussian kernel to compute the trust density as,

$$\rho(x) = \frac{1}{n}\Sigma_i K(Q_{\mathbf{z}'}(\tilde{\mathbf{x}}_i, \mathbf{x}'_i); x) \text{ with } K(Q_{\mathbf{z}'}(\tilde{\mathbf{x}}_i, \mathbf{x}'_i); x) = \frac{1}{\gamma}\sqrt{\frac{n}{2\pi}}\exp\left(-\frac{(Q_{\mathbf{z}'}(\tilde{\mathbf{x}}_i, \mathbf{x}'_i)-x)^2}{2(\gamma/\sqrt{n})^2}\right), \quad (5)$$

where $Q_{\mathbf{z}'}(\tilde{\mathbf{x}}_i, \mathbf{x}'_i)$ is the question-answer trust for each sample (Eq. (4)), $\gamma$ is the kernel parameter, and $n$ is the number of samples in the respective burn depth classes. In this study, $\gamma$ is set to 0.5. This choice of $\gamma$ has been used to compute the trust density of ResNet-50 for image recognition tasks (Wong et al., 2020).

The trust spectrum (Wong et al., 2020) measures overall trust behavior for a given burn depth class $\mathbf{z}'$. It is defined by

$$T_M(\mathbf{z}') = \mathbb{E}[Q_{\mathbf{z}'}]. \quad (6)$$

Here, it is assumed that the probability of the occurrence of question-answer trust has a uniform distribution.

The NetTrustScore (Wong et al., 2020) shows the overall trustworthiness of the model $M$ and it is defined as the expectation of the trust spectrum, *i.e.*

$$S_{net} = \mathbb{E}[T_M(\mathbf{z}')]. \tag{7}$$

## 3. Results

The efficacy of the BurnNet is evaluated in classifying burn depths. First, binary classification of the deep-partial thickness burns vs. the rest of the burn depths is studied. Next, a multiclass classification of four burn depths is examined.

### 3.1. Binary classification

The binary classification is performed to identify deep-partial thickness burn depth (80 samples) from the *in situ* porcine skin dataset comprised of all four burn depths (320 samples). The details of model training are described in Section 2. The performance of the BurnNet is compared with the benchmark classifiers described in Section 2.5. Table 1 shows the confusion matrices for the BurnNet and benchmark classifiers. The confusion matrices were obtained from an independent assessment using 20-fold cross-validation. The null hypotheses $H_0$ indicate that the data point (or burned skin sample) belongs to the deep-partial thickness burn group. Rejecting the null hypothesis suggests that the data point belongs to the rest of the burn group. True and False indicate the actual burn depth class identifier of the samples. From Table 1, we observe that the BurnNet accurately identifies all but two deep-partial burn samples in contrast with other classifiers.

Table 1. Confusion matrices for the binary classification.

| Classifier | Decision | True | False |
|---|---|---|---|
| LDA | Accept $H_0$ | 36 | 23 |
|  | Reject $H_0$ | 44 | 217 |
| SVM | Accept $H_0$ | 50 | 6 |
|  | Reject $H_0$ | 30 | 234 |
| VGG16 | Accept $H_0$ | 49 | 4 |
|  | Reject $H_0$ | 31 | 236 |
| ResNet50 | Accept $H_0$ | 35 | 7 |
|  | Reject $H_0$ | 45 | 233 |
| DenseNet121 | Accept $H_0$ | 47 | 5 |
|  | Reject $H_0$ | 33 | 235 |
| **BurnNet** | **Accept $H_0$** | **78** | **0** |
|  | **Reject $H_0$** | **2** | **240** |

Table 2 gives the performance metrics computed from the confusion matrices (Table 1). From Table 2, we see that with 0.99 accuracy, 0.98 sensitivity, 1.00 specificity, 0.99 F-score, and 0.98 MCC score, the performance metrics for the BurnNet are significantly higher than the other benchmark classifiers. This indicates that the BurnNet effectively characterizes the distinctive texture patterns between the deep-partial thickness burn and the rest of the burn group images. The high specificity for the benchmark classifiers (0.90-0.98) suggests that they capture the texture pattern of the rest of the burn samples. However, their low sensitivity (0.44-0.63) indicates that they do not recognize texture patterns of deep partial-thickness burn samples. This is also manifested in their low F-score (0.60-0.73) and MCC score (0.40-0.69). The high value of MCC score and F-score for the BurnNet further underscores its ability to classify burn depths with high accuracy despite an unbalanced dataset, with the *in situ* porcine skin dataset having three times fewer deep-partial thickness burned than the rest of the burn samples.

Table 2. Performance metrics of the classifiers.

| Classifiers | Accuracy | Sensitivity | Specificity | F-score | MCC score |
|---|---|---|---|---|---|
| LDA | 0.79 | 0.45 | 0.90 | 0.60 | 0.40 |
| SVM | 0.88 | 0.63 | 0.98 | 0.76 | 0.68 |
| VGG16 | 0.89 | 0.61 | 0.98 | 0.75 | 0.69 |
| ResNet50 | 0.83 | 0.44 | 0.97 | 0.60 | 0.52 |
| DenseNet121 | 0.88 | 0.59 | 0.98 | 0.73 | 0.67 |
| **BurnNet** | **0.99** | **0.98** | **1.00** | **0.99** | **0.98** |

We further examine the performance of the classifiers using ROC and PR curves obtained from 20-fold cross validation. Fig. 3(a) compares ROC curve of BurnNet (AUC = 0.99) with that of the benchmark classifiers (AUC between 0.68 to 0.86). The high AUC value indicates that the BurnNet separates the deep-partial thickness burns and the rest of the burns samples with more significant margins than the benchmark classifiers. Fig. 3(b) compares the PR curve of the classifiers. The high AUC value of 0.95 for the PR curve indicates the robustness of the BurnNet even with the unbalanced dataset.

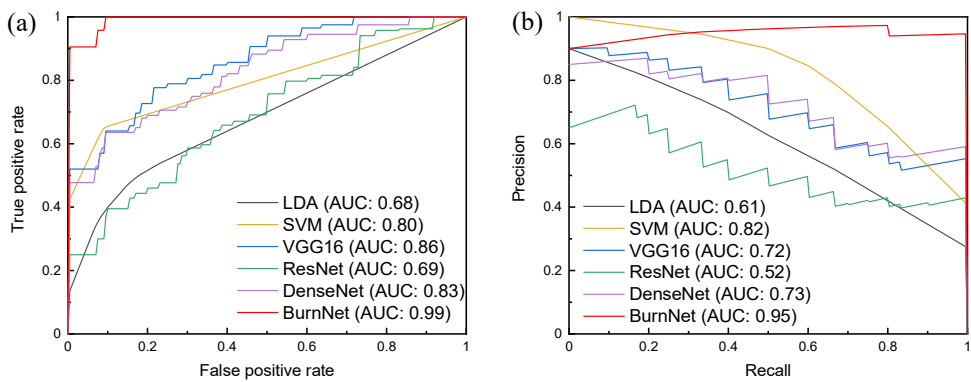

**Fig. 3.** (a) Receiver operating characteristic (ROC) curves, and (b) precision-recall (PR) curves computed from 20-fold cross-validation.

Next, we benchmark the trustworthiness of the BurnNet against well-known classifiers for the binary classification. We compute question-answer trust, trust density, trust spectrum, and NetTrustScore (Wong et al., 2020) as described in Section 2.6 to measure the trustworthiness of the classifiers. Fig. 4 shows the trust density of the BurnNet and benchmark classifiers (iii)-(v). Note that the trust density is not defined for the LDA and SVM models. From Fig. 4, we see that the trust density of the BurnNet is significantly concentrated around the point where question-answer trust is 1. This indicates that the BurnNet classifies the burn depths with high confidence compared to the other benchmark classifiers.

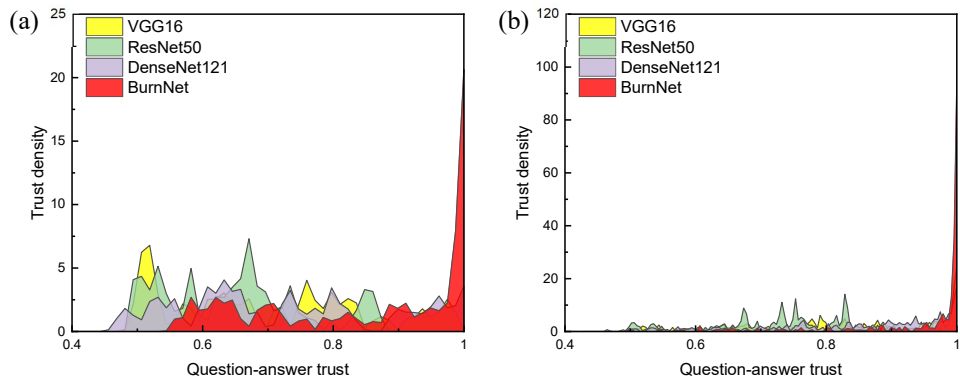

**Fig. 4.** Trust density of the classifiers for the (a) deep-partial thickness burn and (b) and the rest of the burn depths.

Table 3 gives the trust spectrum of deep-partial thickness burns $T_M^{DP}$ and the rest of the burns class $T_M^R$, and NetTrustScore $S_{net}$. From Table 3, we see that trust spectrum of BurnNet is significantly higher than the benchmark classifiers. This tells us that the BurnNet classifies the burn depths with a greater margin, resulting in higher NetTrustScore.

**Table 3.** Trustworthiness metrics of the classifiers.

| Classifiers | $T_M^{DP}$ | $T_M^R$ | $S_{net}$ |
|---|---|---|---|
| VGG16 | 0.70 | 0.83 | 0.76 |
| ResNet50 | 0.65 | 0.74 | 0.69 |
| DenseNet121 | 0.73 | 0.85 | 0.79 |
| **BurnNet** | **0.85** | **0.95** | **0.90** |

### 3.2. Multiclass classification

We evaluate the efficacy of the BurnNet in classifying four depths of tissue injury as described in Section 2.1. Fig. 5 gives the confusion matrix obtained from 20-fold cross validation. The number of samples that are correctly classified is shown along the diagonal in the confusion matrix. The off-diagonal components correspond to the number of misclassified samples. The multiclass classification accuracy is estimated to be 95%. Table 4 gives the trustworthiness metrics for multiclass classification, *i.e.,* trust spectrum of superficial partial burns $T_M^{SP}$, deep partial burns $T_M^{DP}$, light full-thickness burns $T_M^{LFT}$, deep full-thickness burns $T_M^{DFT}$, and NetTrustScore $S_{net}$. Fig. 6 shows the trust density plots of the classifiers for four burn depths. Compared to trust density of the benchmark classifiers, BurnNet shows significantly higher trust density for all four burn depths. These manifest in the high confidence of BurnNet in multiclass classification.

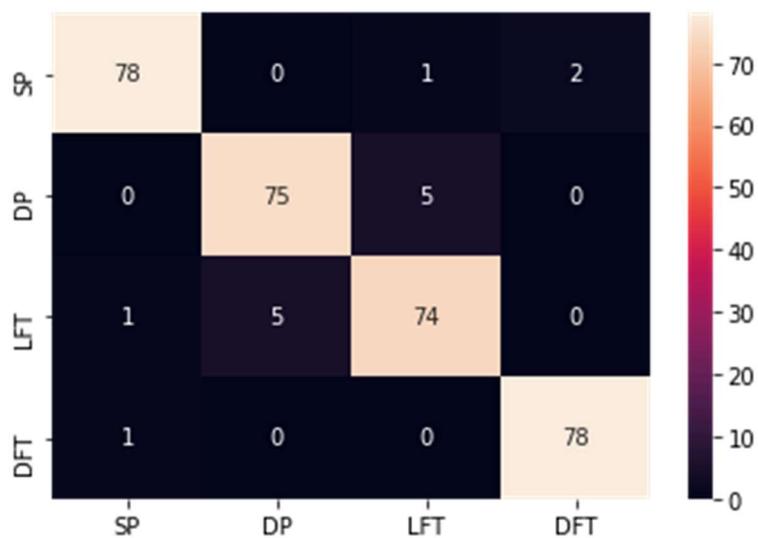

**Fig. 5**. Confusion matrix for multiclass classification. SP: Superficial partial-thickness; DP: Deep partial thickness; LFT: Light full thickness; DFT: Deep full thickness.

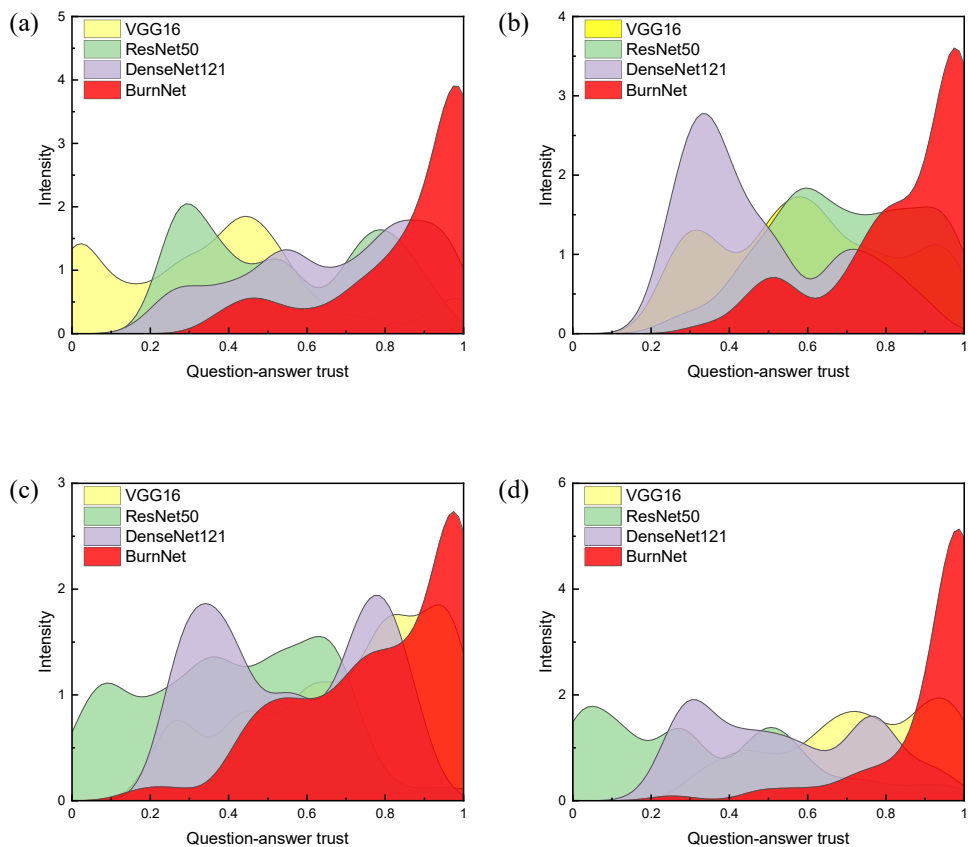

**Fig. 6.** Trust density of the classifiers for the (a) superficial partial thickness burn, (b) deep-partial thickness burn, (c) light full thickness burns and (d) deep full thickness burn.

Table 4. Trustworthiness metrics for multiclass classification.

| Classifier | $T_M^{SP}$ | $T_M^{DP}$ | $T_M^{LFT}$ | $T_M^{DFT}$ | $S_{net}$ |
|---|---|---|---|---|---|
| VGG16 | 0.49 | 0.47 | 0.56 | 0.48 | 0.50 |
| ResNet50 | 0.55 | 0.48 | 0.56 | 0.55 | 0.53 |
| DenseNet121 | 0.67 | 0.60 | 0.69 | 0.73 | 0.67 |
| **BurnNet** | **0.93** | **0.92** | **0.91** | **0.93** | **0.92** |

## 4. Discussion

The performance metrics presented in the previous section elucidate the efficacy of the BurnNet in classifying burn depths. This section examines how the BurnNet localizes the altered textural pattern in the B-mode images of burned skin to classify the burn depths. For this task, we employ a combination of *post hoc* explainability tools – generalized gradient-weighted class activation mapping (Grad-CAM++) in conjunction with guided backpropagation, *i.e.,* guided Grad-CAM++.

Guided GradCam++ provides a visual explanation for the CNN model by localizing target objects in image classification tasks (Chattopadhay et al., 2018). It shows how sensitive the logit is as the convolution layer of interest changes. Guided GradCam++ combines the advantages of guided backpropagation and GradCam++. Guided backpropagation generates images where fine-grained details such as edges and ridges are emphasized (Selvaraju et al., 2017). GradCam++ visualizes class-discriminative image areas; however, it cannot show the fine-grained details of the image. Hence, by performing elementwise multiplication of the guided backpropagation and GradCam++ the guided GradCam++ produces a class-discriminative region of images with fine-grained details (Chattopadhay et al., 2018).

We study guided GradCam++ heatmaps to identify the contributing regions in the B-mode images of four burn depth groups as described in Section 2.1. Fig. 7 shows the guided GradCam++ heatmaps averaged over samples in each of the four burn depth groups. The averaged heatmaps are normalized to [0,1]. From Fig. 7, we see that with increasing burn depth, more pixels are getting activated, particularly along the depth of the ultrasound image. This is clearly reflected in the increased intensity of the heatmaps with increasing depth of burn injuries. The mean and standard deviation of heatmap intensities along the depth of the image shown in Fig. 8 further highlights the increase in activation level with increasing burn severity. The heatmap of deep full-thickness burns (Fig. 7(d)) shows a relatively high activation of pixels at the top of the image, where burn treatment is applied to the skin.

These findings are consistent with our observation from Fig. 1, which shows increased speckle density along the depth of the ultrasound images with increasing depth of burns. The speckles are the manifestation of microstructural degradation of thermally damaged skin tissue. Thermal damage induces contrasting acoustic impedance in the burned skin tissues (Fujii et al., 2012). The strong scattering signals from the burn site with contrasting acoustic impedance produce intricate interference patterns that appear as speckles in the ultrasound images. Hence, the burn conditions directly influence the speckle pattern in ultrasound images, which is the basis of burn depth classification. BurnNet utilizes this variation in the speckle patterns to classify burn depth.

The guided GradCam++ heatmaps elucidate BurnNet's ability to localize the speckle pattern alteration across the ultrasound images and the area with significant damage to the dermal layers, such as the case with full-thickness burned samples (Lee et al., 2020). This is crucial for burn depth classification, clearly evident from a significant improvement in performance metrics for BurnNet over benchmark classifiers.

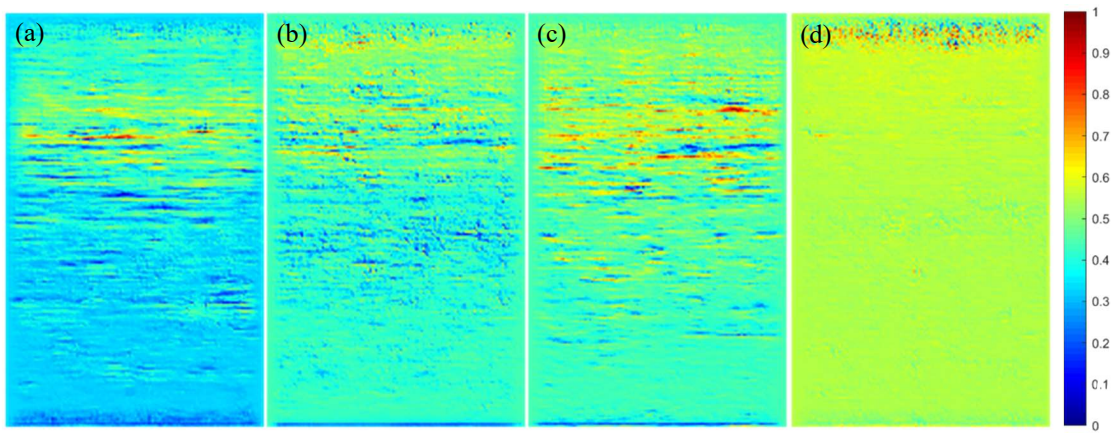

**Fig. 7.** Guided GradCam++ heatmaps averaged over samples in each of the four burn depths, *i.e.,* (a) superficial-partial thickness, (b) deep-partial thickness, (c) light full-thickness, and (d) deep full-thickness burns. The color bar indicates the intensity of guided GradCam++ activation.

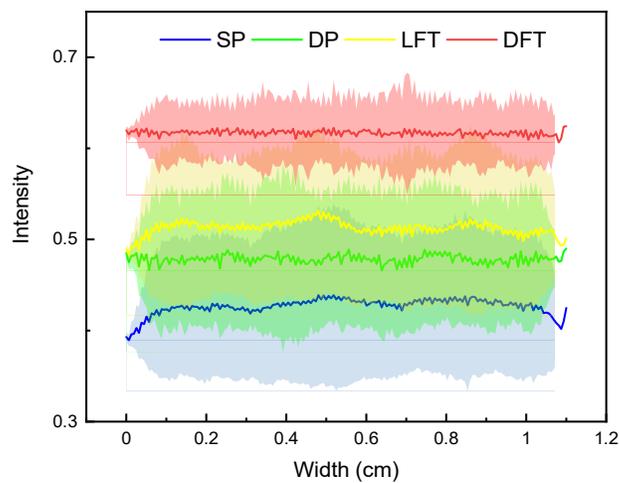

**Fig. 8.** Mean and standard deviation of heatmap intensities (Fig. 7) along the depth of the ultrasound image. SP: Superficial partial thickness; DP: Deep partial thickness; LFT: Light full thickness; DFT: Deep full thickness.

## 5. Conclusion

A deep convolutional neural network architecture is presented for the classification of burn depths using diagnostic ultrasound images. A *post hoc* explanation of the network elucidates its ability to capture alterations in the textural pattern with increasing depth of burn injuries, which is crucial for burn classification. This is evident from a significant improvement in the performance metrics of the network over state-of-the-art benchmark classifiers. The proposed network is shown to identify deep-partial thickness burns, which are most challenging to diagnose clinically, in a binary classification task with significantly higher accuracy, specificity, and sensitivity of 99%, 100%, and 98%, respectively, compared to classifiers used in the benchmark study. A high value of AUC for the PR curve (0.95) and corresponding F-score (0.99), and MCC score (0.98) further indicates the robustness of the network despite the unbalance dataset used in this study. A multiclass classification study shows that the network can classify four burn depths with 95% accuracy. The proposed network has demonstrated excellent performance metrics for burn depth classification on *in situ* porcine skin datasets. However, for clinical applications, it needs to be tested on *in vivo* datasets, which may introduce additional uncertainties in textural patterns associated with perfusion and tissue echogenicity.

## Acknowledgments

The authors gratefully acknowledge the support of this work through the U.S. Army Futures Command, Combat Capabilities Development Command Soldier Center STTC cooperative research agreement #W911NF-17-2-0022.